\documentclass[
    aps,12pt,notitlepage,
    oneside,onecolumn,nofootinbib,
    nolongbibliography,centertags
    aps,amsmath,amssymb,a4paper
]{revtex4-2}

\usepackage{graphicx}
\graphicspath{{figures/}}

\usepackage[usenames,dvipsnames]{xcolor} % link colors tweak

\usepackage[
    colorlinks = true,
    urlcolor = MidnightBlue,
    linkcolor = Red,
    citecolor = NavyBlue
]{hyperref}

\hypersetup{
    pdfauthor = {A. V. Glushkov, L. T. Ksenofontov, K. G. Lebedev, A. Saburov},
    pdftitle = {Muon puzzle in ultra-high energy EAS according to...},
    pdfsubject = {extensive air showers},
    pdfkeywords = {ultra-high energy cosmic rays, extensive air showers, muon puzzle, the Yakutsk array, the Pierre Auger Observatory}
}

\usepackage{array}
\newcolumntype{a}[1]{>{\centering\arraybackslash\hspace{0pt}}p{#1}}

\def\FigWidth{0.65\textwidth}

\newcommand{\fig}[1]{fig.~\ref{#1}}
\newcommand{\eq}[1]{(\ref{#1})}

\newcommand{\yeas}{YEASA}

\newcommand{\DOI}[2] {%
    \href{https://doi.org/#1}{#2}%
}
\newcommand{\arXiv}[2] {%
    arXiv:~\href{https://arXiv.org/abs/#1}{#1} [#2]%
}
\newcommand{\arXivOld}[2] {%
    arXiv:~\href{https://arXiv.org/abs/#2/#1}{#2/#1}%
}

\newcommand{\E}{E_0}

\newcommand{\rhosSOOT}{\rho_{\text{SD}}(600,\theta)}

\newcommand{\rhomSOOT}{\rho_{\text{MD}}(600,\theta)}

\newcommand{\meancos}{\left<\cos\theta\right>}

\newcommand{\Ebin}{E_{\text{bin}}}

\newcommand{\Ethin}{E_{\text{thin}}}
\newcommand{\wmax}{w_{\text{max}}}

\newcommand{\lnA}{\langle\ln{A}\rangle} % <lnA>
\newcommand{\xmax}{x_{\text{max}}}      % Xmax

\newcommand{\RhoExp}{\rho_{\text{MD}}^{\text{exp}}}
\newcommand{\RhoP}{\rho_{\text{MD}}^p}
\newcommand{\RhoFe}{\rho_{\text{MD}}^{\text{Fe}}}
\newcommand{\RhoPAO}{\rho_{\text{MD}}(450, 35\degr)}
\newcommand{\RhoPAOT}{\rho_{\text{MD}}(450,\theta)}
\newcommand{\RhosYkt}{\rho_{\text{SD}}(600, 25.8\degr)}
\newcommand{\RhosYktO}{\rho_{\text{SD}}(600, 0\degr)}
\newcommand{\RhomYkt}{\rho_{\text{MD}}(600, 25.8\degr)}

\newcommand{\qgs}{{\sc qgsj}et01}
\newcommand{\qgsii}{{\sc qgsj}et-{\sc ii}.04}

\newcommand{\eposlhc}{{\sc epos-lhc}}

\newcommand{\fluka}{{\sc fluka}2011}
\newcommand{\corsika}{{\sc corsika}}

\newcommand{\sqrm}{~m$^2$}     % m^2
\newcommand{\degr}{^{\circ}}       % degrees

\begin{document}

\title{Muon puzzle in ultra-high energy EASs according to Yakutsk array and Auger experiment data}

\author{A.\,V.\,Glushkov\textsuperscript{1,}}
\email{glushkov@ikfia.ysn.ru}

\author{A.\,V.\,Saburov}
\email{vs.tema@gmail.com}

\author{L.\,T.\,Ksenofontov\textsuperscript{1,}}
\email{ksenofon@ikfia.ysn.ru}

\author{K.\,G.\,Lebedev\textsuperscript{1,}}
\email{LebedevKG@ikfia.ysn.ru}

\affiliation{\textsuperscript{$\rm 1$}Yu.\,G.\,Shafer Institute of Cosmophysical Reserach and Aeronomy of Siberian branch of the Russian Academy of Sciences,\\
31 Lenin ave., Yakutsk, 677027, Russia}

\begin{abstract}
    The lateral distribution of particles in extensive air showers from cosmic rays with energy above $10^{17}$~eV registered at the Yakutsk complex array was analyzed. Experimentally measured particle densities were compared to the predictions obtained within frameworks of three ultra-high energy hadron interaction models. The cosmic ray mass composition estimated by the readings of surface-based and underground detectors of the array is consistent with results based on the Cherenkov light lateral distribution data. A comparison was made with the results of direct measurement of the muon component performed at the Pierre Auger Observatory. It is demonstrated that the densities of muon flux measured at Yakutsk array are consistent with results of fluorescent light measurements and disagree with results on muons obtained at the Auger array.
\end{abstract}

\maketitle

\section{Introduction}

The problem of increased muon content in extensive air showers (EAS) from ultra-hight energy cosmic rays (UHECR) in comparison to model predictions has been noted by researchers for more than 20 years~\cite{bib:1}. In the combined analysis published by international working group on this problem the data of eight EAS arrays have been considered: EAS-MSU, IceCube, KASCADE-Grande, NEVOD-DECOR, The Pierre Auger Observatory (Auger), SUGAR, Telescope Array (TA) and Yakutsk complex EAS array~\cite{bib:2}. For comparison of data from different experiments a scaling parameter was introduced:

\begin{equation}
    z = \frac{
        \ln\RhoExp - \ln\RhoP
    }{
        \ln\RhoFe - \ln\RhoP
    }\text{,}
    \label{eq:1}
\end{equation}

\noindent
where $\RhoExp$ is muon density measured in experiment; $\RhoP$ and $\RhoFe$~--- are muon densities calculated in simulated showers from primary protons ($p$) and iron nuclei (Fe) as recorded by detectors of a certain array. As a result it was shown that model calculations agree with experiment up to $10^{16}$~eV. However this situation changes with further increase of primary energy. A wide spread of $z$ values is observed, especially within ultra-high energy domain in strongly inclined EASs~\cite{bib:3} and at large distances from shower axis~\cite{bib:4}. In the case of Yakutsk array muon densities determined at axis distance 300~m in showers with primary energy $\E \ge 10^{18}$~eV and average zenith arrival direction $\meancos = 0.9$ were used. In the case of \qgs~\cite{bib:5} model they gave the value of $z$ parameter $z = 0$ and for \qgsii~\cite{bib:6} and \eposlhc~\cite{bib:7}~--- negative values~\cite{bib:2}. In work~\cite{bib:8} the muon fraction was investigated at axis distances 300, 600 and 1000~m in showers with $\E \simeq 10^{17.7 - 19.5}$~eV and $\meancos = 0.9$. In works~\cite{bib:9, bib:10} zenith-angular dependencies of the muon fraction were considered at 600~m from the axis in showers with $\E \simeq 10^{18}$ and $10^{19}$~eV and with zenith directions $\cos\theta \ge 0.5$. All these works~\cite{bib:8, bib:9, bib:10} confirm the agreement between the experiment and predictions of \qgs{} model for primary protons ($z \simeq 0$).

\begin{figure}[!htb]
    \centering
    \includegraphics[width=\FigWidth]{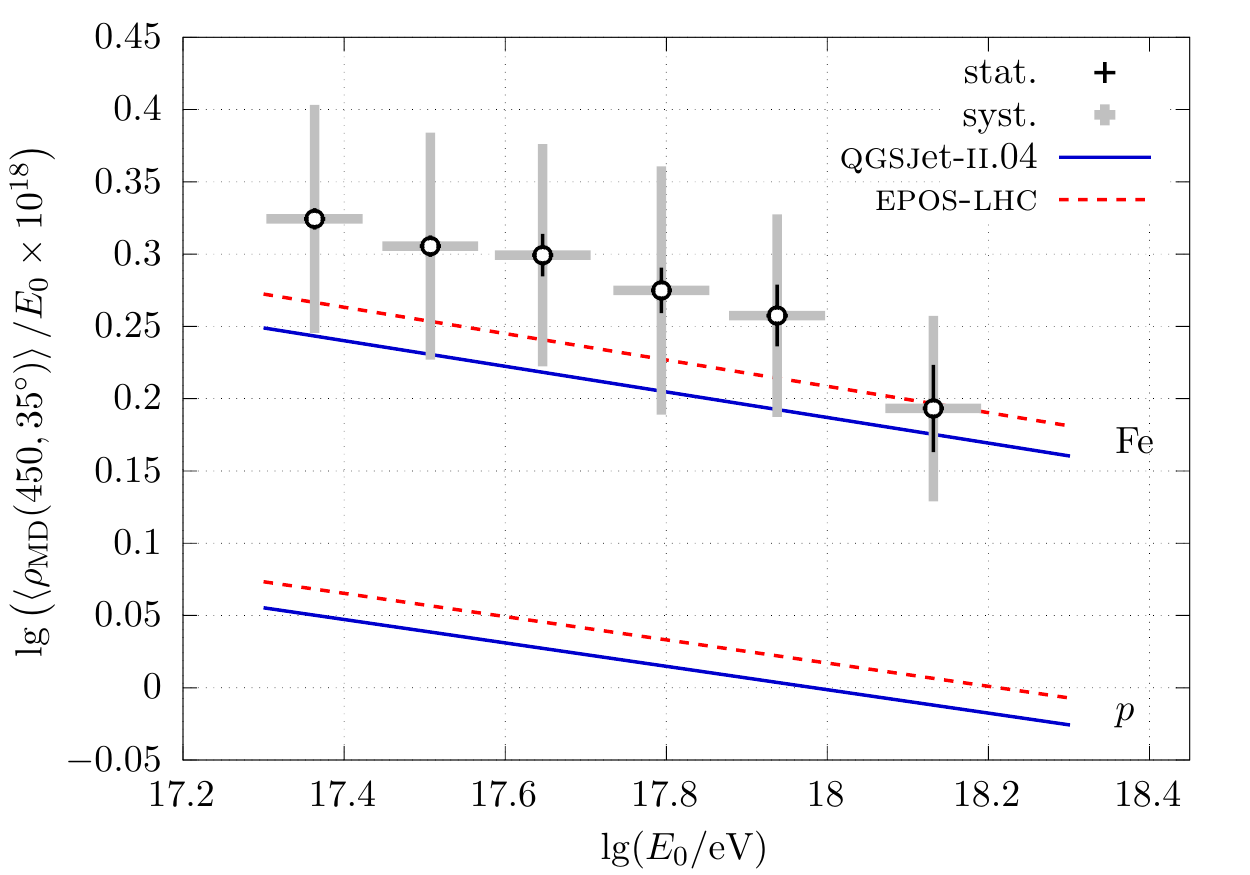}
    \caption{Muon densities in EAS at axis distance 450~m normalized by primary energy. Results of the direct measurements performed at the Auger array with underground scintillation detectors with $\simeq 1.0 \times \sec{35\degr}$~GeV threshold. The data were taken from fig.~11 in work~\cite{bib:11}.}
    \label{fig:1}
\end{figure}

The Auger collaboration have reported on direct measurements of muons in EAS with energies $2 \times 10^{17}-2 \times 10^{18}$~eV and zenith angles $\theta \le 45\degr$~\cite{bib:11}. The measurements were performed with 5 and 10-\sqrm{} scintillation detectors with registration threshold $\simeq 1.0 \times \sec\theta$~GeV (placed under a 2.3-m layer of ground). One of these result is presented on \fig{fig:1}. In this work the $\RhoPAO$ parameters was considered~--- muon density measured in individual showers at axis distance 450~m converted to zenith angle $35\degr$ with following relations:

\begin{gather}
    \RhoPAO =
    \frac{
        \RhoPAOT
    }{
        f_{\text{att}}(\theta)
    }\text{,}
    \label{eq:2} \\
    f_{\text{att}}(\theta) =
    1 + (0.54 \pm 0.10) \cdot x + (1.02 \pm 0.69) \cdot x^2\text{,}
    \label{eq:3}
\end{gather}

\noindent
where $x = \cos^2\theta - \cos^2{35\degr}$. It is worth mentioning that in this experiment at Auger array, the muon component of EAS was separated directly, similar to how it was done at Yakutsk array. Direct interpretation of the results presented in~\cite{bib:11} does not exclude the possibility that the considered events have originated from primary iron nuclei. These results contradict not only the conclusions of works~\cite{bib:8, bib:9, bib:10}, but also measurements made in the very same experiment but with different technique.

\begin{figure}[!htb]
    \centering
    \includegraphics[width=\FigWidth]{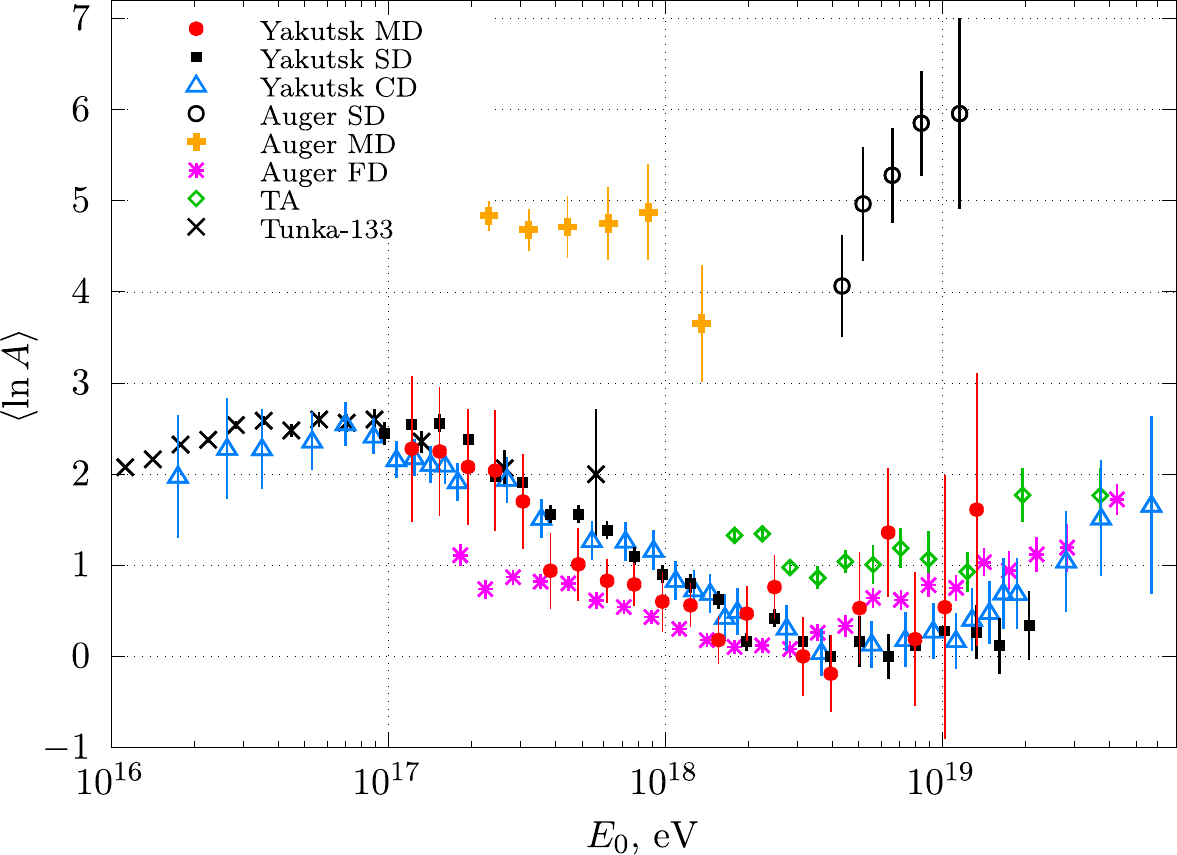}
    \caption{Estimations of the mean CR mass composition following from data of several experiments within the framework of \qgsii{} model. Results of Yakutsk array were obtained with three independent techniques~\cite{bib:12, bib:13, bib:14}. Estimations for Auger array were calculated from the values of $z$ parameter obtained within the framework of \qgsii{} model for muon data (Auger MD), data of stations constituting the surface trigger (Auger SD) and for results of the $\left<\xmax\right>$ measurements (Auger FD)~\cite{bib:11} (see also \fig{fig:5}). The data of Tunka-133~\cite{bib:15} and TA~\cite{bib:16} are also shown.}
    \label{fig:2}
\end{figure}

On \fig{fig:2} estimations of the cosmic ray (CR) mass composition are presented, obtained in several experiments with different techniques within the framework of the \qgsii{} model. The displayed results of Yakutsk array were obtained with three independent methods: from lateral distribution function (LDF) of muon component registered with muon detectors (MD) with $\simeq 1$~GeV threshold~\cite{bib:12}; from the shape of LDF of charged and electromagnetic components registered with surface-based detectors constituting the main trigger of the array (SD)~\cite{bib:13}; and by measuring the flux of Cherenkov light (CL) emitted by EAS (CD)~\cite{bib:14}. It is seen that all three component of EAS give results that are consistent with each other within experimental errors. They are also consistent with estimations obtained from CL data at Tunka-133 array~\cite{bib:15} and with values calculated from average maximum depth of EAS cascade curves ($\left<\xmax\right>$) measured at TA~\cite{bib:16}. Estimations that follow from values of $z$ parameter obtained at the Auger array~\cite{bib:11} are represented by three sets of data: according to $\left<\xmax\right>$ measurements made by registering the fluorescent light emission of EAS (FD), measurements of muon component with $\simeq 1$~GeV threshold (MD) and surface component in strongly inclined showers (SD). As it follows from \fig{fig:2}, the direct measurements of muon component with underground scintillation detectors in inclined showers at the Auger array stand out from the general trend and encourage us to consider them in more detail.

In this article we analyze the data obtained during the long-standing period of the Yakutsk complex EAS array (\yeas) operation. We compare them with the Auger results~\cite{bib:11}. They are directly related to each other: both experiments use similar scintillation detectors for muon registration which are calibrated with similar techniques, by cosmic muon background. This allowed a direct comparison between the data of two arrays.

\section{Lateral distribution of EAS particles according to the Yakutsk array data}

In works~\cite{bib:17, bib:18} responses of the \yeas{} surface-based and underground scintillation detectors were calculated in showers initiated by CRs with energy above $10^{17}$~eV. With the use of \corsika{} code~\cite{bib:19} a set of artificial air showers was generated in energy range $10^{17} - 10^{19.5}$~eV with logarithmic step $\Delta\lg(\E/\text{eV}) = 0.5$ and with zenith angles $0\degr-60\degr$. Simulations were performed with the use of ultra-high energy interaction models \qgs~\cite{bib:5}, \qgsii~\cite{bib:6} and \eposlhc~\cite{bib:7}. Hadron interactions at energies below 80~GeV were treated with \fluka{} code~\cite{bib:20}. To speed-up the calculation, the thin-sampling mechanism was utilized~\cite{bib:20} with thinning level $\Ethin = (10^{-6}-10^{-5})$ and weight limit for all components $\wmax = \E \cdot \Ethin$. 200 events were simulated for each set of primary parameters $(\E, \theta)$. Based on the statistics of each set, mean LDFs (MLDF) of detector response was calculated with radial logarithmic binning of axis distance with $\Delta\lg(r/\text{m}) = 0.04$ step. Energy dependencies of the \yeas{} surface-based and underground detectors responses to shower particles are presented on \fig{fig:3}. The values were obtained for the axis distance 600~m within frameworks of three hadron interaction models. All particle densities were converted to primary energy $10^{19}$~eV by multiplying by the normalization ratio $10^{19}/\E$. Average densities derived from simulation results were compared to the values obtained from experimental data with the MLDF method.

\subsection*{Obtained results}

For the analysis, showers were selected with axes located within a 1-km circle around the array center and determined with an accuracy no worse than 50~m (see Table~\ref{t:1}). The set of selected events was divided into energy intervals (bins) with $\Delta\lg(\E/\text{eV}) = 0.2$ step. In each bin an MLDF of particle density was calculated~--- for particles detected with surface-based detectors (SD) and underground detectors with $\simeq 1$~GeV threshold (MD). From the resulting MLDFs particle densities were determined at axis distance 600~m~--- $\RhosYkt$ and $\RhomYkt$. The accuracy of estimated densities was no worse than 10\%. Showers energy was estimated by the formula:

\begin{equation}
    \E = (3.76 \pm 0.30) \times 10^{17}
    \cdot
    \RhosYktO^{1.02 \pm 0.02}~\text{[eV],}
    \label{eq:4}
\end{equation}

\noindent
where $\RhosYktO$ is EAS classification parameter $\RhosYkt$ converted to vertical arrival direction~\cite{bib:22}. During the construction of MLDF the recorded particle densities of both components were multiplied by the normalization ratio $\Ebin / \E$, where $\Ebin$ is average energy within a bin.

\begin{figure}[!htb]
    \centering
    \includegraphics[width=0.66\textwidth]{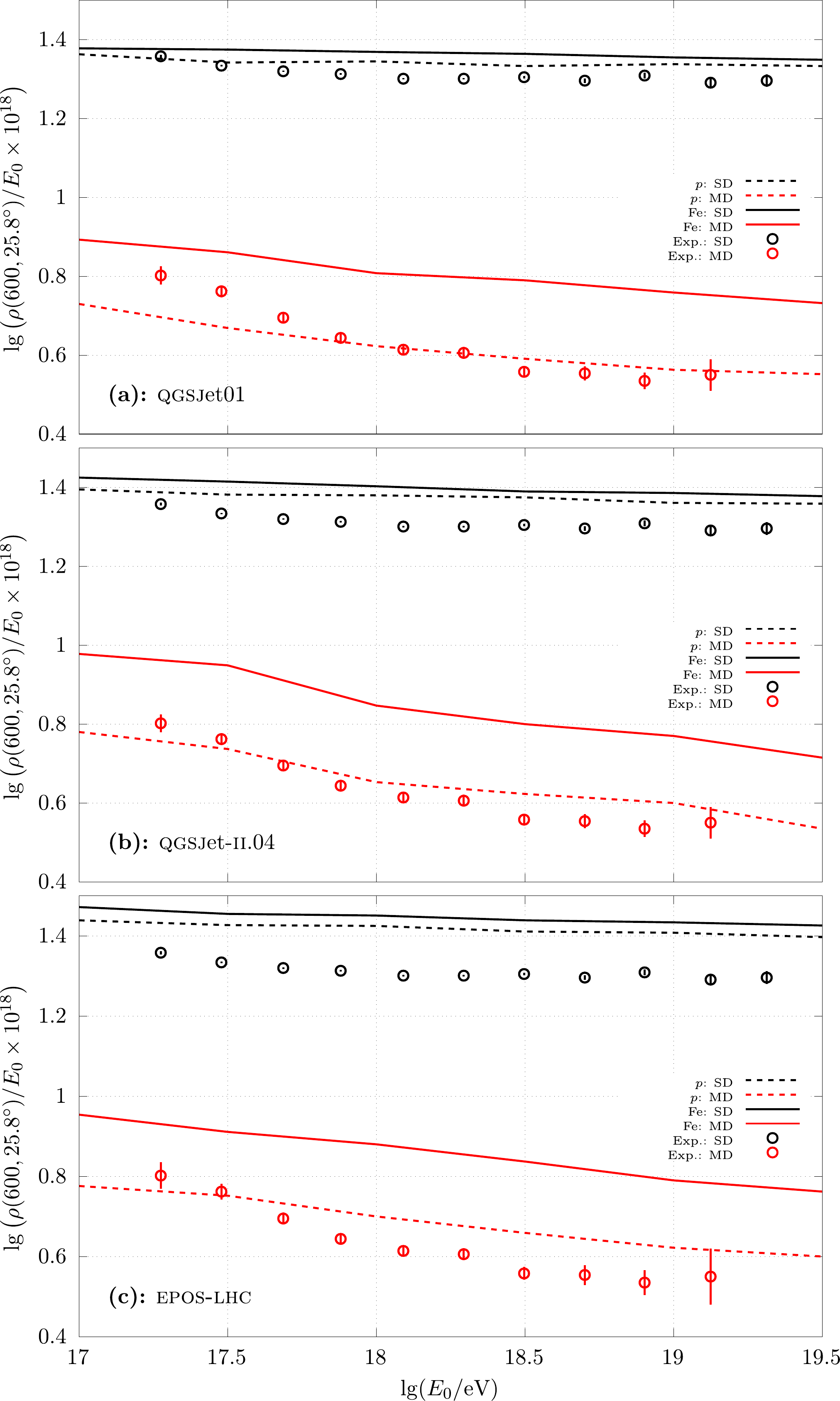}
    \caption{Averaged responses of surface-based (SD) and underground scintillation detectors with $1.0 \times \sec\theta$~GeV threshold (MD) at 600~m from the axis normalized by primary energy of EAS with average zenith direction $\meancos = 0.9$. Lines~--- results of simulations performed within frameworks of three hadron interaction models for primary protons ($p$) and iron nuclei (Fe). Symbols~--- average values obtained from experimental data.}
    \label{fig:3}
\end{figure}

\begin{table*}[htb]
    \centering
    %\caption{Таблица\,1. Число ливней в выборках СФПР. $N_{\text{sh}}$~--- число событий, вошедших в бин с данной средней энергией.}
    \caption{Number of showers in MLDF samples. $N_{\text{sh}}$ is the number of events within a bin with a certain average energy.}
    \label{t:1}
    \begin{tabular}{%
        c
        a{0.075\textwidth}a{0.075\textwidth}a{0.075\textwidth}
        a{0.075\textwidth}a{0.075\textwidth}a{0.075\textwidth}
        a{0.075\textwidth}a{0.075\textwidth}a{0.075\textwidth}
        a{0.075\textwidth}
    }
        \hline
        $\left<\lg(\E/\text{eV})\right>$ &
        17.28 & 17.48 & 17.68 & 17.88 & 18.09 &
        18.29 & 18.49 & 18.70 & 18.90 & 19.12 \\
        \hline
        $N_{\text{sh}}$ &
        6079 & 6182 & 4807 & 2717 & 1316 &
        600 & 260 & 107 & 60 & 16 \\
        \hline
    \end{tabular}
\end{table*}

It is seen on \fig{fig:3} that responses of surface-based and underground detectors from EAS particles turned out to be lower than expected from primary protons, and muon densities~--- significantly lower. This result is possible due to various reasons. One of them can be connected with energy estimation in experiment. First factor in equation \eq{eq:4} reflects the systematic error of 8\% arising from the very method of calibration adopted at Yakutsk array~\cite{bib:22}. To understand the above mentioned result, let's assume that energy of showers presented on \fig{fig:3} was overestimated by the amount of disagreement between theory and experiment for surface-based detectors. In the case of \qgs{} model, to reach agreement between densities measured with surface-based detectors and those obtained in simulation it is sufficient to reduce the proportionality ratio in equation \eq{eq:4} by $\sim 10\%$. Muon densities should also rise by $\simeq 10\%$ after re-normalization of energy. In such a case, at $\E > 10^{18}$~eV both measured EAS components would agree with simulation results within experimental errors. In energy region below $10^{18}$~eV muon densities rise higher and higher as energy decreases. This can be interpreted as a change in composition of primary particles due to addition of a certain fraction of heavier nuclei to protons. A similar behaviour of primary particles composition is observed on \fig{fig:1} (change towards heavier nuclei with decrease of energy) but against the background of iron nuclei. The surface component of EAS on \fig{fig:3} shows similar tendency but not so pronounced due to its weaker dependence on muons. The \qgsii{} model gives similar result with reduction of proportionality ratio in expression \eq{eq:4} by $\simeq 15$\% and \eposlhc~--- with reduction by $\simeq 20$\%.

\begin{figure}[htb]
    \centering
    \includegraphics[width=\FigWidth]{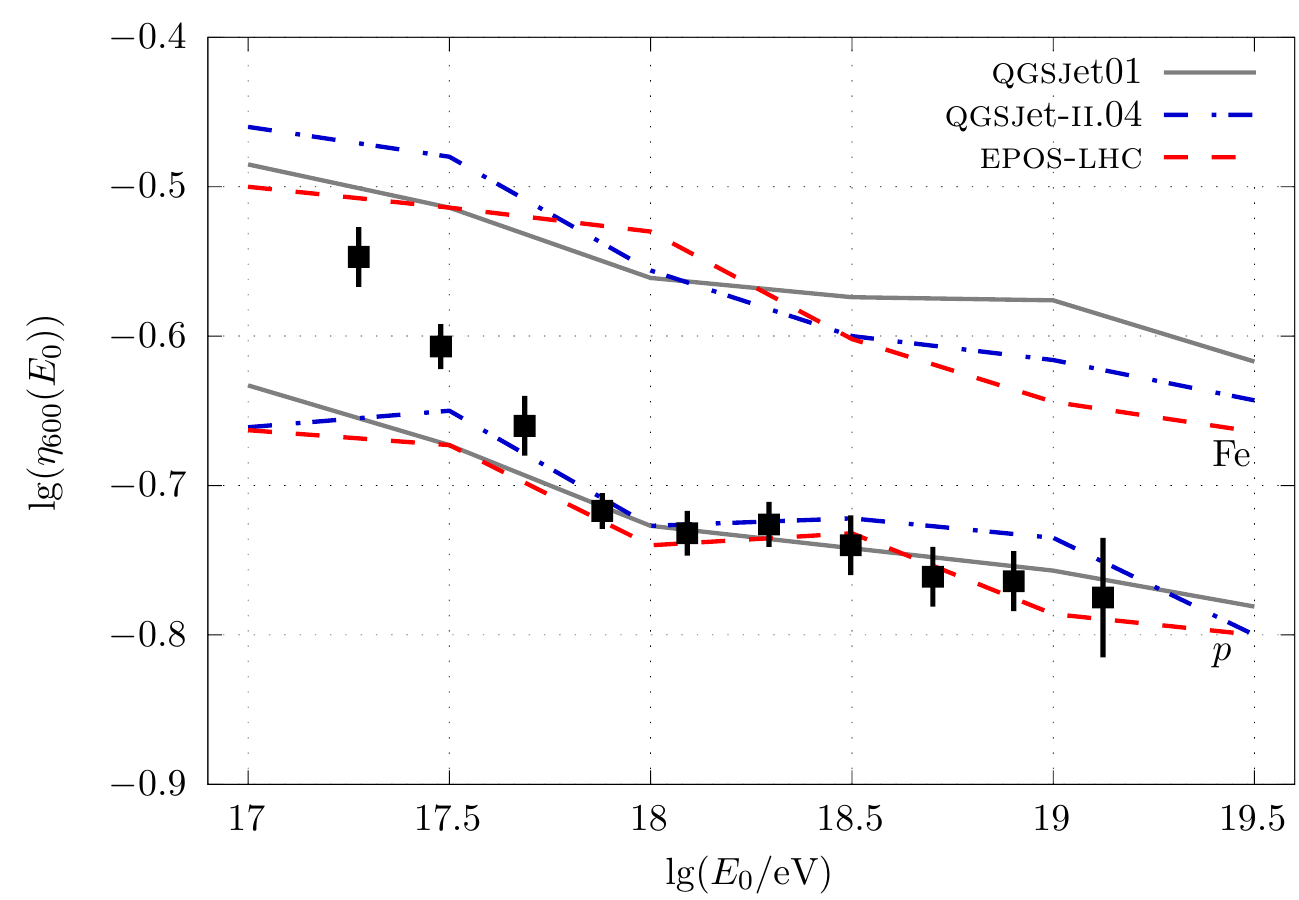}
    \caption{Energy dependence of the muon fraction in EASs with average zenith direction $\meancos = 0.9$ at 600~m from the axis. Symbols denote the values obtained from experimental data, lines~--- theoretical predictions obtained within frameworks of three hadron interaction models for primary protons ($p$) and iron nuclei (Fe).}
    \label{fig:4}
\end{figure}

At first glance the above-mentioned results are critically sensitive to primary energy. However, since MLDFs of both EAS components on \fig{fig:3} were obtained from the shared sample of events with average energy $\Ebin$, the muon fraction

\begin{equation}
    \eta_{600}(\E) =
    \frac{
        \left<\rho_{\text{MD}}(600)\right>/\Ebin
    }{
        \left<\rho_{\text{SD}}(600)\right>/\Ebin
    } = \frac{
        \left<\rho_{\text{MD}}(600)\right>
    }{
        \left<\rho_{\text{SD}}(600)\right>
    }
    \label{eq:5}
\end{equation}

\noindent
does not depend on energy. On \fig{fig:4} the muon fraction $\eta_{600}(\E)$ is shown which was obtained from the data presented on \fig{fig:3}.

Simulations demonstrate that the value of $\eta_{600}$ reflects the physical meaning of $z$ parameter \eq{eq:1}:

\begin{equation}
    z = \frac{
        \ln{\eta^{\text{exp}}_{600}} - \ln{\eta^p_{600}}
    }{
        \ln{\eta^{\text{Fe}}_{600}} - \ln{\eta^p_{600}}
    }\text{.}
    \label{eq:6}
\end{equation}

The values of $z$ parameter obtained this way for different energy intervals $\Ebin$ are listed in Table~\ref{t:2}. All three models at $\E \ge 10^{17}$~eV within measurement errors do not contradict the hypothesis of a pure proton composition of primary cosmic rays (i.e. $z \simeq 0$).

\begin{table*}[htb]
    \centering
    %\caption{Таблица\,2. Значения параметра $z$, вычисленные из доли мюонов на расстоянии от оси ШАЛ 600~м (\fig{fig:4}) с помощью соотношения \eq{eq:6}. В колонке $\delta{z}$ приведены ошибки, учитывающие как систематические, так и статистические неопределенности.}
    \caption{Values of the $z$ parameter obtained from muon fraction at EAS axis distance 600~m (\fig{fig:4}) with the use of the relation \eq{eq:6}. The $\delta{z}$ column lists errors that include both systematic and statistical uncertainties.}
    \label{t:2}
    \begin{tabular}{%
        a{0.13\textwidth}%
        a{0.13\textwidth}a{0.13\textwidth}a{0.13\textwidth}%
        a{0.13\textwidth}a{0.13\textwidth}a{0.13\textwidth}%
    }
        \hline
        $\lg(\Ebin / \text{eV})$ &
        \multicolumn{2}{c}{\qgs} &
        \multicolumn{2}{c}{\qgsii} &
        \multicolumn{2}{c}{\eposlhc} \\
        \hline
        &
        $z$ & $\delta{z}$ &
        $z$ & $\delta{z}$ &
        $z$ & $\delta{z}$ \\
        \hline
        17.28 &  0.69 & 0.13 &  0.58 & 0.11 &  0.76 & 0.12 \\
        17.48 &  0.41 & 0.09 &  0.25 & 0.09 &  0.41 & 0.09 \\
        17.68 &  0.19 & 0.12 &  0.11 & 0.12 &  0.21 & 0.12 \\
        17.88 &  0.00 & 0.07 & -0.04 & 0.07 &  0.04 & 0.07 \\
        18.09 &  0.00 & 0.09 & -0.01 & 0.09 &  0.02 & 0.09 \\
        18.29 &  0.06 & 0.09 &  0.00 & 0.09 &  0.06 & 0.09 \\
        18.49 &  0.01 & 0.11 & -0.06 & 0.16 & -0.06 & 0.15 \\
        18.70 & -0.07 & 0.11 & -0.10 & 0.17 & -0.06 & 0.14 \\
        18.90 & -0.05 & 0.11 & -0.06 & 0.13 &  0.07 & 0.14 \\
        19.12 & -0.10 & 0.22 & -0.06 & 0.29 &  0.08 & 0.28 \\
        \hline
    \end{tabular}
\end{table*}

The errors include both events statistics during the construction of mean LDFs and all other errors arising during events processing (calibration of detectors, reconstruction of arrival direction and axis coordinates, estimation of energy etc.). They are difficult to separate, and it is not necessary. They are accumulated in average values $\rhosSOOT$ and $\rhomSOOT$ (see e.g., \fig{fig:3}).

\begin{figure}[htb]
    \centering
    \includegraphics[width=\FigWidth]{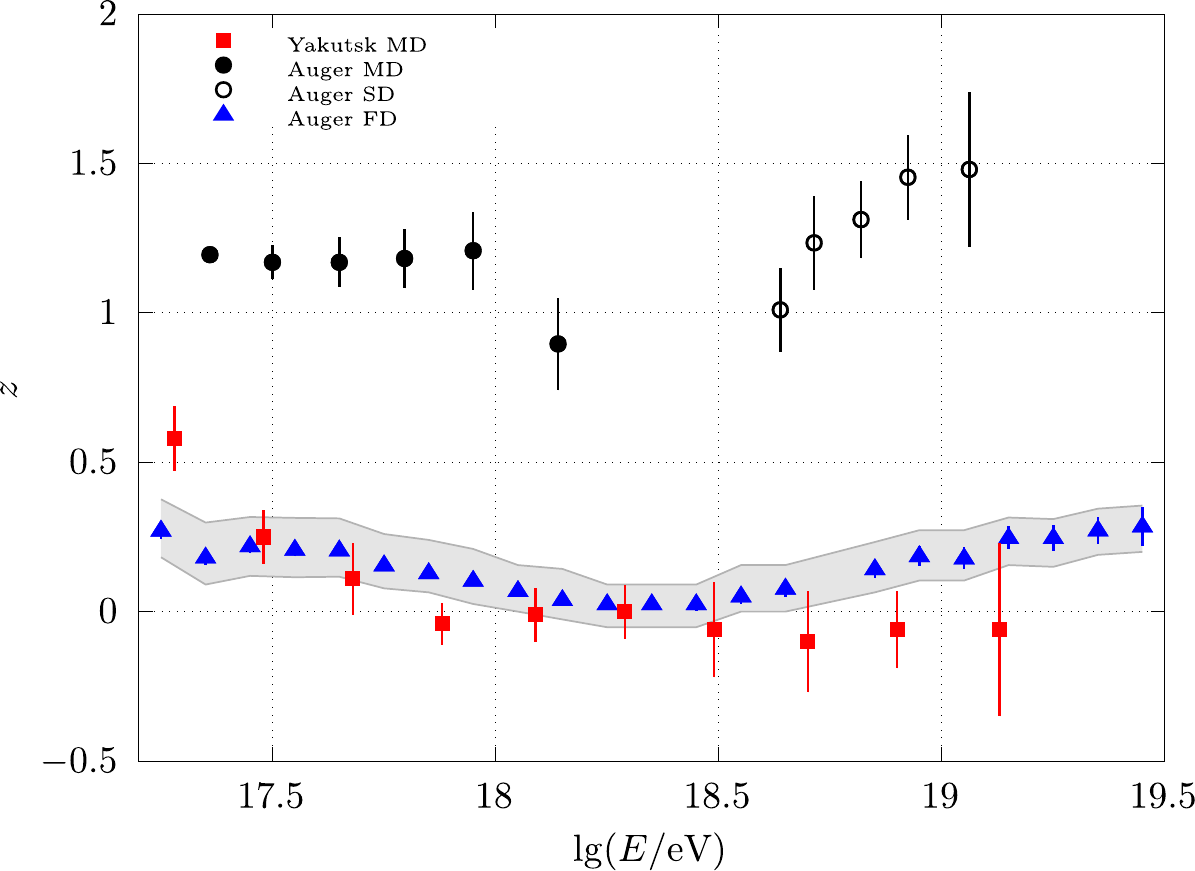}
    \caption{Energy dependencies of the $z$ parameter according to different measurement techniques at Yakutsk array and Auger Observatory obtained within the framework of \qgsii{} model. The Auger data are represented with direct measurements of muon component of EAS (Auger MD), the analysis of particle densities measured with surface detectors at 1000~m from the axis (SD) and with estimations obtained from direct observations of shower maximum with optical method (Auger FD). The gray band denotes systematic uncertainties of optical measurements at Auger. The Yakutsk data are listed in Table~\ref{t:2} (Yakutsk MD).}
    \label{fig:5}
\end{figure}

The $z$ parameter is a part of a simple and important relation

\begin{equation}
    \left<\ln{A}\right> = z \cdot \ln{56}\text{.}
    \label{eq:7}
\end{equation}

\noindent
which is used for estimation of the mean atomic weight $A$ of primary nuclei. Its connection with muons only speaks of the fact that muons are sensitive to CR composition. But not only muons. Estimations of the CR composition obtained from other components of EAS are shown on \fig{fig:2}. A comparison of values of the parameter $z$ obtained at Auger and \yeas{} within the framework of \qgsii{} model is shown on \fig{fig:5}. The Auger data are represented with values obtained from particle densities measured with muon detectors (Auger MD) and surface detectors (Auger SD) in inclined showers at axis distance 1000~m. They are consistent with each other and indicate abnormal muon densities. Also on \fig{fig:5} are presented values obtained from measurements of the $\left<\xmax\right>$ (Auger FD). These data were extracted from fig.~13{\sl b} of the work~\cite{bib:11}. They do not contradict the expected light composition of primary particles which is close to pure protons. The values of $z$ parameter obtained at Yakutsk array are listed in fourth and fifth columns of Table~\ref{t:2}. The values of $\lnA$ calculated with the expression \eq{eq:7} from the Auger data presented on \fig{fig:5} (Auger MD, Auger SD and Auger FD) are shown on \fig{fig:2}.

\section{Comparison of data of two arrays and discussion}

\begin{figure}[htb]
    \centering
    \includegraphics[width=\FigWidth]{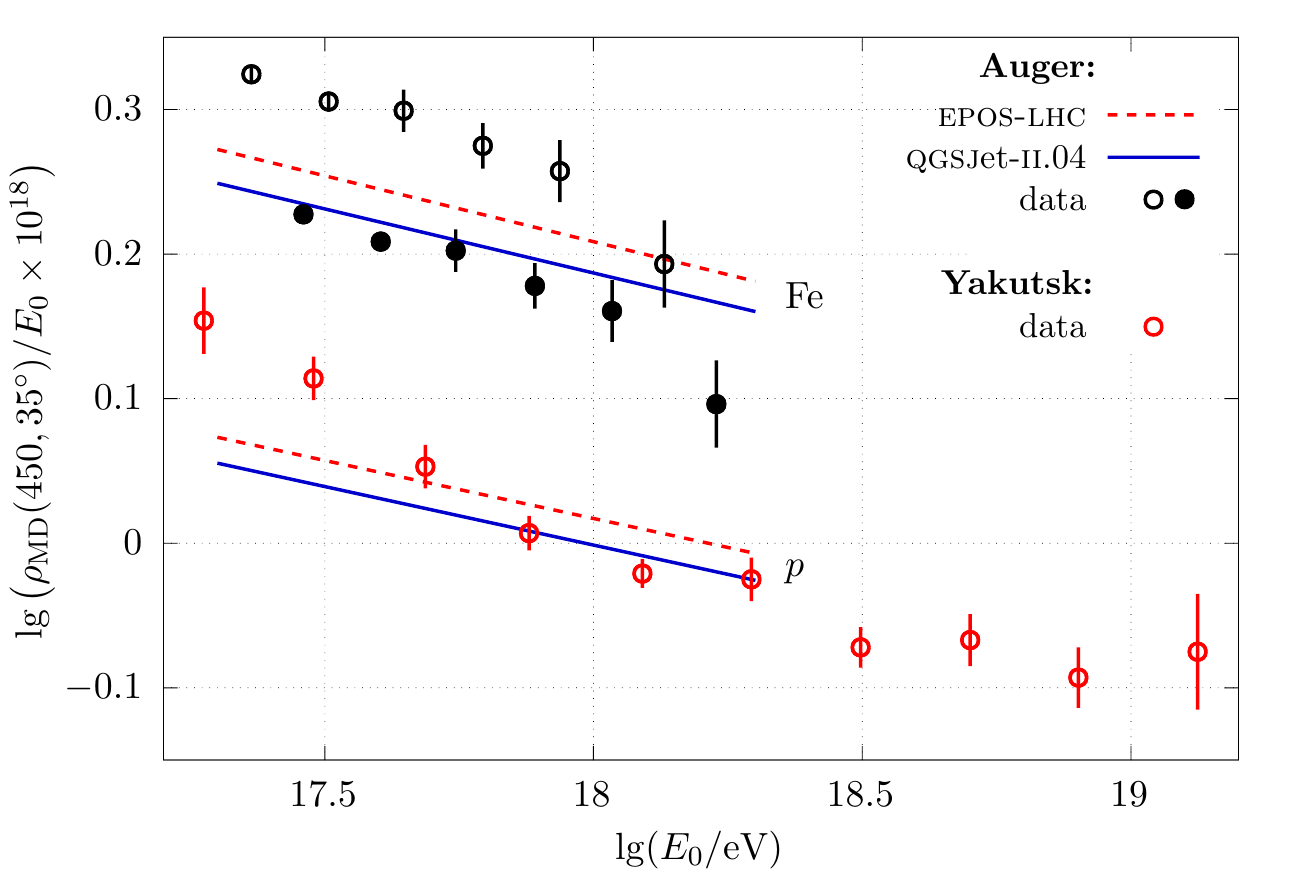}
    \caption{Comparison of energy dependencies of the $\RhoPAO$ parameter obtained by Auger Collaboration (see fig.~11 in work~\cite{bib:11}) and by \yeas. Average muon densities obtained in Yakutsk experiment were converted to zenith angle $\theta = 35\degr$. Lines represent the results of simulation performed for muon detectors of the Auger array within frameworks of \qgsii{} and \eposlhc{} models for primary protons ($p$) and iron nuclei (Fe). Symbols denote experimental data. Dark circles represent muon densities renormalized after the Auger energy estimation was increased by factor 1.25.}
    \label{fig:6}
\end{figure}

In work~\cite{bib:11} to register the muon component of EAS the Auger Collaboration used scintillation detectors with $1.0 \times \sec\theta$~GeV threshold~--- similar to Yakutsk experiment. Both arrays calibrate their muon detectors by cosmic muon background. This makes it possible to directly compare these experiments with each other. For this comparison we have selected muon densities recorded at axis distance 450~m ($\rho_{\text{MD}}(450,25.8\degr)$) from the Yakutsk events sample. With the use of expressions \eq{eq:2} and \eq{eq:3} they were converted to the $\RhoPAO$ value which was used in measurements at the Auger. The resulting values are shown on \fig{fig:6} which essentially is \fig{fig:1} with superimposed Yakutsk data. At $\E \ge 8 \times 10^{17}$~eV they are consistent with simulation results performed for Auger detectors within the framework of \qgsii{} model for primary protons, and at lower energies point at mixed composition of primary particles.

The recently heavily discussed problem of abnormally high muon content in EASs of ultra-high energies (see e.g. \cite{bib:1, bib:2, bib:3, bib:4, bib:11, bib:13}) is not entirely conditioned by muons. Many experiments that measure the muon component in practice normalize muon densities to primary energy. In the pair ``muons--primary energy'' the latter is the weak link. Muon density divided by primary energy is a dimension value (m$^{-2}$/eV). The value of primary energy directly effects the results (whether ``many'' of ``few'' muons were observed) compared to other experiments and model calculations. Although the actual number of muons could be relatively normal.

In work \cite{bib:22} primary CR energy spectra obtained by the Auger Collaboration and \yeas{} are compared. The agreement between both spectra is quite possible if, for example, one increases the Auger primary energy estimation by factor 1.25. As a result of such a re-normalization of the Auger data, all densities displayed on \fig{fig:6} should be reduced by 25\% (denoted with dark circles). The values obtained with this recalibration may well be related to heavy composition of primary particles. In such a case the muon puzzle (i.e. the discrepancy between theory and experiment) loses all its urgency and enters the mainstream of a constructive search for the sources of the remaining disagreements.

\section{Conclusion}

The muon fraction obtained in a combined analysis of particle densities $\left<\rho_{\text{MD}}(600, 25.8\degr)\right>$ and $\left<\rho_{\text{SD}}(600, 25.8\degr)\right>$ on the total sample of air showers with energies from $2 \times 10^{17}$~eV up to $2 \times 10^{19}$~eV is consistent with the expected values obtained within frameworks of \qgs, \qgsii{} and \eposlhc{} hadron interaction models (see \fig{fig:4}). The comparison of these data with the results of The Auger Collaboration~\cite{bib:11} have shown that they directly contradict each other (see \fig{fig:5} and \fig{fig:6}). Furthermore, in contrast to heavily discussed ``muon excess'' in other experiments, the results of Yakutsk array at higher energies starts to look more like ``muon deficit'' compared to model predictions. At $\E \ge 8 \times 10^{17}$~eV the hypothesis of a pure proton cosmic ray composition is quite plausible. In the domain of lower energies the composition of primary particles apparently is mixed, with addition of heavy nuclei. This is consistent with our earlier estimations~\cite{bib:12, bib:13, bib:24, bib:25}. The work~\cite{bib:11} draws attention to the internal contradiction of the results which, in our opinion, could be conditioned by some peculiarities of the Auger experiment. It is obvious that the problem of measuring the muon EAS component requires further detailed investigation.

\acknowledgements

This work was made within the framework of the state assignment No. 122011800084-7 using the data obtained at The Unique Scientific Facility ``The D. D. Krasilnikov Yakutsk Complex EAS Array'' (\url{https://ckp-rf.ru/catalog/usu/73611/}). Authors express their gratitude to the staff of the Separate structural unit \yeas{} of ShICRA SB RAS.

\end{document}